%% file: main.tex
\documentclass[aps,prl,twocolumn,superscriptaddress,groupedaddress]{revtex4}

\usepackage[utf8]{inputenc}
\usepackage[T1]{fontenc}

\usepackage{amsmath}
\usepackage{graphicx}
\usepackage{filecontents}

\begin{document}

\title{Interplay of Excitonic Complexes in $p$-Doped WSe$_2$ Monolayers}

\input author_list.tex

\date{\today}

\begin{abstract}
WSe$_2$ monolayers with variable doping are an ideal system to study two-dimensional excitonic complexes. Here, we find that the strongest photoluminescence from bright neutral excitons occurs at moderate $p$-doping levels, owing to a rapid decrease of signal from positive dark trions and a slow increase of signal from positive bright trions with growing $p$-concentrations. We explain our observations with a qualitative model, in which the scattering rate of bright excitons into dark complexes is enhanced by exciton localization, while the scattering rate into positive bright trions increases with $p$-doping level.

\end{abstract}

\maketitle

As a consequence of their direct band gap \cite{PhysRevLett.105.136805, Splendiani2010} and large exciton binding energies \cite{Ugeda2014, PhysRevLett.113.026803, PhysRevLett.113.076802, PhysRevLett.114.097403}, transition metal dichalcogenide monolayers (TMD-MLs) have become the material of choice for the study of excitonic physics in the two-dimensional limit \cite{RevModPhys.90.021001, Mak2016}. The large spin-orbit coupling, combined with a broken inversion symmetry, results in unique electronic structures, giving rise to the concept of valleytronics \cite{PhysRevLett.108.196802, Schaibley2016}. Furthermore, their purely two-dimensional nature, as well as their mechanical properties, enable strain \cite{He2013, Lloyd2016, Castellanos-Gomez2013, Palacios-Berraquero2017, Manzeli2017} and dielectric \cite{Raja2017, PhysRevMaterials.1.054001, Steinleitner2018} engineering. 

The number of optically observable excitonic complexes is especially high in W-based TMD-MLs, due to their spin-inverted band gap at the K-points of the hexagonal Brillouin zone \cite{Korm_nyos_2015, PhysRevB.93.121107}. The portfolio of identified excitonic complexes in W-based MLs includes neutral bright excitons, charged bright excitons (or trions) \cite{PhysRevB.96.085302}, neutral and charged biexcitons \cite{You2015, Paur2019, Ye2018, Barbone2018, Li2018}, grey trions \cite{Danovich2017, Tu_2019}, as well as spin-forbidden dark excitons \cite{wang2017, Zhang2017} and trions \cite{PhysRevLett.123.027401, Li2019} and momentum-indirect dark excitons \cite{1904.04711}. Exciton-polarons have been reported from highly doped MoSe$_2$ \cite{Sidler2017,PhysRevB.95.035417}. 

The spin-forbidden dark states have attracted much interest due to their long lifetimes \cite{PhysRevB.96.155423, PhysRevLett.115.257403, Zhang2017}. These states can be brightened intrinsically through a spin-orbit interaction induced band mixing \cite{wang2017} or extrinsically through magnetic field induced spin mixing \cite{Zhang2017, Molas_2017} or near-field coupling to surface plasmon-polaritons \cite{Zhou2017}.

Due to the relatively low spin-orbit coupling in the conduction band of W-based TMD-MLs, the largest variety of charged excitonic complexes is observed in electron-doped MLs. The formation of negative trions is very efficient since their binding energy equals the A$^{^\prime}_{1}$ optical phonon energy \cite{Jones2015, Tu_2019}. As a consequence, virtual negative trions appear in the photoluminescence spectra even at low electron concentrations \cite{PhysRevLett.122.217401}. Due to the stronger spin-orbit coupling of the valence band, hole-doped MLs host fewer charged excitonic complexes and, as a result, attracted less attention.

In this work, we scrutinize the interplay of excitonic complexes in hole-doped WSe$_2$ MLs. In gate-dependent $\mu$-photoluminescence experiments, we observe excitonic behaviour, which cannot be explained through the band structure of perfect WSe$_2$ MLs alone. The observed phenomena are qualitatively explained with a phenomenological microscopic model of exciton localization. 

\begin{figure}
    \centering
    \includegraphics{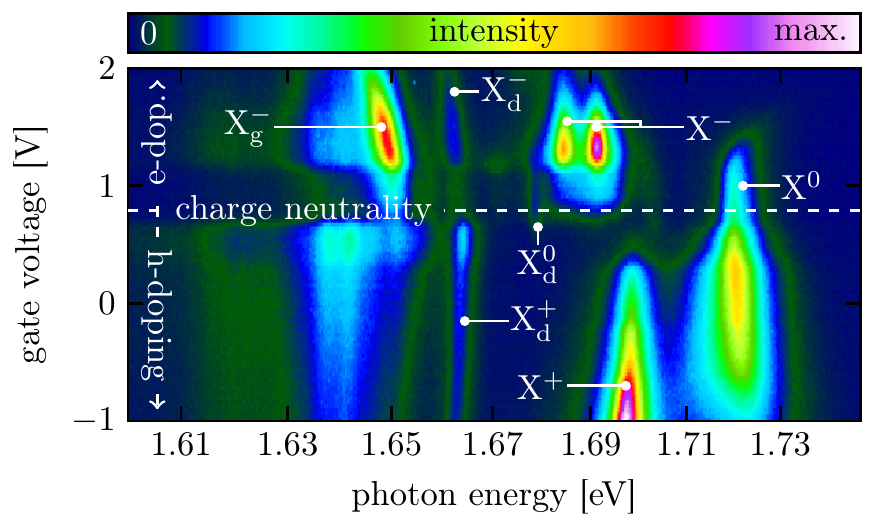}
    \caption{Gate-dependent photoluminescence signal of a WSe$_2$ ML. At the upper (lower) end of the plot, the WSe$_2$ ML is highly electron(hole)-doped. The charge neutrality point ($0.79~\mathrm{V}$) is defined as the gate voltage, at which the combined intensity of all trion complexes is minimal. 
    }
    \label{fig:161310_gate_dependence}
\end{figure}

Figure \ref{fig:161310_gate_dependence} shows low-temperature gate-dependent photoluminescence spectra of a WSe$_2$ ML displayed as a color map (see SI for details on methods, a discussion on the charge density induced in the ML, as well as photoluminescence spectra at selected gate voltages). The highest energy emission ($1.721~\mathrm{eV}$), which is present in a wide range of gate voltages, is attributed to the radiative recombination of the neutral bright exciton (X$^0$). The signal from the neutral dark exciton (X$^0_\mathrm{d}$, $1.678~\mathrm{eV}$) is observable only in a narrow voltage range around $0.9~\mathrm{V}$. In the electron-doped regime ($V_g > 0.7~\mathrm{V}$), the most prominent excitonic features are the fine-structure split negative trion (X$^-$, $1.684~\mathrm{eV}$ and $1.691~\mathrm{eV}$), the negative dark trion (X$_\mathrm{d}^-$, $1.662~\mathrm{eV}$), as well as X$_\mathrm{g}^-$ ($1.649~\mathrm{eV}$), which originates from the recombination of an inter-valley exciton interacting with an extra electron \cite{Danovich2017, Tu_2019}. In the hole-doped regime ($V_g < 0.7~\mathrm{V}$), the spectra contain emission from the positive bright trion (X$^+$, $1.698~\mathrm{eV}$), the positive dark trion (X$_\mathrm{d}^+$, $1.663~\mathrm{eV}$) and a broad band at low energy ($1.640~\mathrm{eV}$). The radially polarized signal beams from the dark excitonic states clearly irrefutably confirms their nature \cite{2001.08043}. Note that the low intensity of the dark states' photoluminescence cannot be used as a measure of their population, due to the different radiation patterns of dark and bright excitonic complexes \cite{wang2017}. 

It is common standard to define the charge neutrality point as the voltage at which the emission from the neutral bright exciton is strongest. This method is not applicable to WSe$_2$ MLs, as will be discussed later. A better definition of the charge neutrality point is given by the voltage, which minimizes the collective trion emission. Using this definition, an inherent $p$-doping of the ML can be seen as the charge neutrality point is shifted to a positive gate bias. The emission energies, linewidths, and gate-dependencies of the observed excitonic complexes agree with recent works on gate-tunable WSe$_2$ MLs \cite{Paur2019, Ye2018, Barbone2018, Tu_2019, Li2018, PhysRevLett.123.027401,Li2019}.

In the electron-doped regime, the photoluminescence intensities of bright and dark excitonic states follow each other when the doping level is changed. The neutral bright and dark excitons are present up to a gate voltage of approximately $1.15~\mathrm{V}$. Around this voltage, the emission lines from negative bright and dark trions intensify and reach their peak intensities in unison at approximately $1.35~\mathrm{V}$. In contrast to the $n$-doped regime, there is a noticeable difference in the evolution of the intensities of bright and dark excitonic complexes in the $p$-doped regime. While the emission from the neutral dark exciton quickly vanishes as holes are being added to the system, the neutral bright exciton emission increases in intensity and reaches its maximum intensity at about $-0.1~\mathrm{V}$. Even more interestingly, the bright and dark trions in the $p$-type regime show opposing evolutions with the intensity of the positive bright (dark) trion emission increasing (decreasing) as a function of hole concentration. 

\begin{figure}
    \centering
    \includegraphics{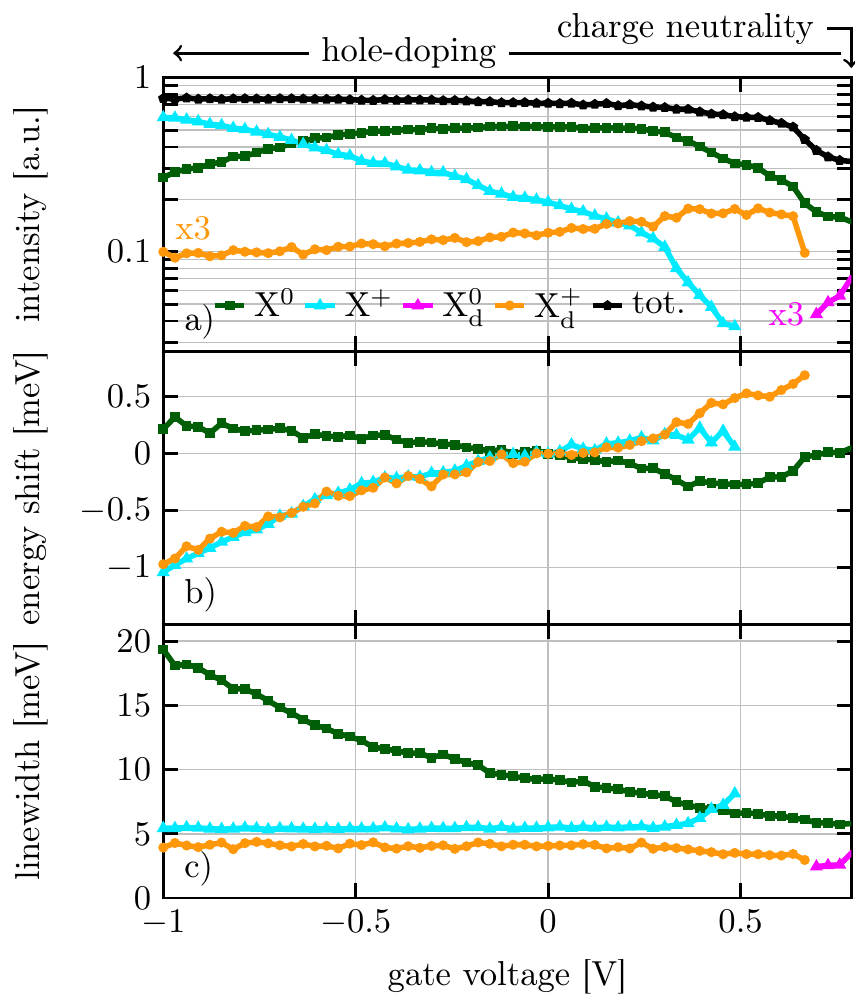}
    \caption{Fitted parameters for excitonic complexes in the hole-doped WSe$_2$ ML. a) Integrated intensity as a function of the applied gate voltage shown on a semi-log scale. b) Energy shift with respect to the energy at $0$~V gate voltage. c) Linewidth.}
    \label{fig:161310_gate_dependence_2}
\end{figure}

In order to analyze the data further, each spectrum in the $p$-type range was fitted using a set of Voigt profiles, each of them corresponding to a different excitonic line. Figure \ref{fig:161310_gate_dependence_2} summarizes the fitted parameters for the most prominent excitonic complexes in the $p$-doped regime. It becomes evident that the contributions to the total intensity change gradually as a function of gate voltage (fig.~\ref{fig:161310_gate_dependence_2} a)). At the charge neutrality point, the X$^0$ line dominates the spectra but is relatively weak compared to lower gate voltages. Its intensity increases as holes are being added to the ML and reaches a plateau between $0.25~\mathrm{V}$ and $-0.5~\mathrm{V}$ before dropping again with a further decrease of the gate voltage. The positive bright trion appears in the spectrum at approximately $0.5$~V, and its emission intensity increases with increasing hole density making it the most dominant emission line at approximately $-0.65~\mathrm{V}$. Eventually, the X$^+$ line intensity saturates (SI, fig.~5). In contrast to the X$^+$ line, the intensity of the X$^+_\mathrm{d}$ line shows a maximum at $0.5~\mathrm{V}$ but then decreases with an increasing hole concentration before eventually saturating.

The emission energy of the X$^0$ line reaches a minimum at low hole concentration and then increases gradually when the gate voltage is further decreased (fig.~\ref{fig:161310_gate_dependence_2} b)); at gate voltages below $-1~\mathrm{V}$, it starts to decrease (SI, fig.~5). The emission energies of the X$^+$ and X$^+_\mathrm{d}$ lines, in contrast, decrease monotonically with a decreasing gate voltage. 

The changes in intensity and emission energy of the X$^0$ line are accompanied by a steady increase in its linewidth with increasing $p$-doping (fig.~\ref{fig:161310_gate_dependence_2} c)). In contrast, the linewidths of X$^+$ and X$_\mathrm{d}^+$ change only for the lowest hole concentrations ($V \gtrsim 0.4~\mathrm{V}$) and then saturate. From the exact line shapes of the emission lines (SI, fig.~4), it can be seen that -- at low hole doping -- there is only a minor contribution from inhomogeneous broadening to the absolute linewidth of the excitonic complexes.

The differences between the $n$- and $p$-doped regimes cannot be explained by the difference in the spin-orbit interactions in the conduction and the valence bands or by the potentially non-linear change of the carrier concentration as a function of gate voltage. As elaborated above, bright and dark excitonic complexes and, in particular, bright and dark trions show completely different behaviors in the $p$-doped regime. This peculiar interplay of excitonic complexes requires consideration of non-intrinsic effects. Here, we develop a microscopic model that takes into account localization and doping-dependent scattering rates and fully describes the observed phenomena in a qualitative way.

\begin{figure}
    \centering
    \includegraphics{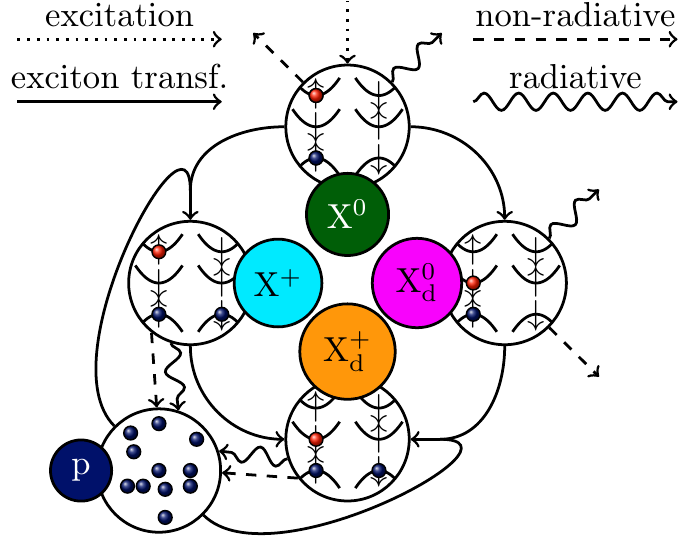}
    \caption{Schematic diagram of transformation paths for excitonic complexes in a $p$-doped WSe$_2$ ML. Neutral bright excitons are generated via optical excitation (dotted line). Solid lines represent the transition from one excitonic complex to another one. Free holes are involved in the formation of bright and dark trions. Curved and dashed lines represent radiative and non-radiative recombination processes, respectively. 
    }
    \label{fig:rate_equation_system}
\end{figure}

Neutral bright excitons are generated at a constant rate, after which they relax to the K-points of the hexagonal Brillouin zone \cite{Steinleitner2017}. Figure \ref{fig:rate_equation_system} shows possible transformations of the excitonic complexes in a $p$-doped WSe$_2$ ML that follow the relaxation; a mathematical description of this model is shown in the supplementary information. Neutral excitons can recombine radiatively or non-radiatively, bind with a  hole into a positive bright trion or be spin-scattered into a neutral dark exciton. Positive bright trions can recombine radiatively or non-radiatively or be spin- or momentum-scattered into positive dark trions. Neutral dark excitons can recombine or bind with a hole into a positive dark trion. Non-radiative recombination of the excitonic complexes shortens their lifetimes and reduces photoluminescence intensities but is otherwise not included in the discussion. The formation of momentum-indirect excitons is included in the non-radiative recombination processes. It is noteworthy, however, that a set of emission lines at low energy (approximately $1.640~\mathrm{eV}$), which have been attributed to the phonon-assisted emission from momentum-indirect excitons, strictly follows the intensity evolution of the positive dark trion line \cite{1904.04711,PhysRevResearch.1.032007}.

\begin{figure}
    \centering
    \includegraphics{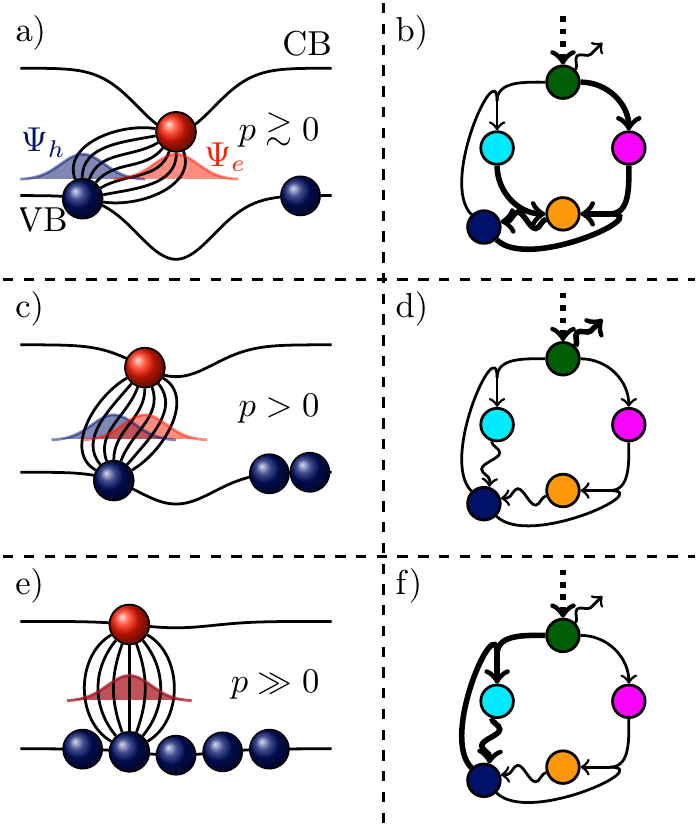}
    \caption{Flattening of the valence and conduction bands (VB and CB) due to hole-induced screening. a), b) and c) show the potential landscape in a WSe$_2$ ML for low, moderate, and high hole doping, respectively. The potential fluctuations are increasingly screened when the density of free holes in the system is increased. The spatial overlap of the wave functions of the hole and the electron forming a neutral bright exciton (blue and red curve, respectively) increases as a function of hole doping. b), d), and f) show the transformation system (fig.~\ref{fig:rate_equation_system}) for the situations depicted in a), b), and c), respectively. The linewidth of the arrows represents the strength of a process.}
    \label{fig:theory_1}
\end{figure}

The probabilities of the transformation and recombination processes depicted in figure \ref{fig:rate_equation_system} may vary as a function of the hole concentration in the system. For a perfect semiconductor, the formation probabilities of neutral complexes remain unchanged when the hole density in the system is increased, whereas the formation probabilities of positively charged excitonic complexes are expected to scale with the hole density. In this case, the highest intensity of X$^0$ would be expected at the charge neutrality point (SI, fig.~6). Furthermore, it would be expected that X$^+$ and X$^+_\mathrm{d}$ show similar intensity trends (SI, fig.~6). The evolution of several photoluminescence lines in figures \ref{fig:161310_gate_dependence} and \ref{fig:161310_gate_dependence_2} is inconsistent with this simple picture and, therefore, points to non-trivial variations of the transition probabilities between different excitonic complexes, as well as those of exciton recombination processes, as a function of doping.

An explanation for the observed behavior can be given assuming spatial fluctuations of the conduction and valence band edge energies of the WSe$_2$ ML and considering the role of a hole gas as energy absorber in the formation process of positive bright trions. Fluctuations in the band edge energies can be generated, for instance, by charged impurities that can be present in the ML itself or its close proximity or through dielectric environment fluctuations \cite{arXiv:1904.04959,Raja2019}. In the supplementary information, we present low-energy emission lines, which suggest the presence of charged impurities in the ML. The importance of the hole gas as energy absorber in the formation process of positive bright trions results from the fact that their binding energy lies within an energy gap where no phonons with suitable symmetry are available \cite{PhysRevB.87.165409}, which is the case in the $n$-doped regime \cite{Jones2015, Tu_2019}.

At a low doping level, the potential fluctuations are unscreened (fig.~\ref{fig:theory_1} a) and b)) and a hole and an electron forming a neutral bright exciton are separated in space. As a consequence, the oscillator strength of this state is decreased and the radiative recombination of the exciton is less likely. At the same time, the formation rate of positive bright trions is low for two reasons. Firstly, the localization of the neutral bright excitons and the low density of free holes result in a low likelihood for these particles to overlap in space. Secondly, the energy released in the formation process of positive bright trions is absorbed very inefficiently by the low number of free holes in the system. As a consequence, the transition to a neutral dark exciton is the most likely channel for neutral bright excitons, whereas only a low number of positive bright trions is formed. Since the neutral dark exciton has a lifetime much longer than the neutral bright exciton \cite{PhysRevLett.115.257403,Zhang2017}, it can bind with a free hole from the low density of holes more efficiently, forming a positive dark trion which eventually recombines radiatively. The low number of positive bright trions can scatter efficiently into positive dark trions, due to the localization-induced relaxation of the momentum mismatch.

When the hole doping level is increased to a moderate level the potential fluctuations are gradually screened (fig.~\ref{fig:theory_1} c) and d)). This increases the wave function overlap between the hole and the electron in a neutral bright exciton resulting in an increase of the oscillator strength of the neutral bright exciton and an associated higher radiative decay rate from this state, which translates into a more intense emission line for X$^0$. As the radiative recombination rate for neutral bright excitons increases, less neutral dark excitons and, consecutively, less positive dark trions are formed, resulting in a weaker X$_\mathrm{d}^+$ signal. At the same time, the increased density of holes in the system leads to the onset of the formation of positive bright trions, as the contact interaction of neutral bright excitons with free holes becomes more likely and the binding energy can be transferred to the hole gas. The fact, that the positive dark trion intensity decreases while the positive bright trion intensity increases rapidly indicates that the radiative decay of the latter is more efficient than the scattering into positive dark trions, which is another consequence of the screening of disorder. 


When the hole doping level is increased further at around $-1.0$~V, the potential fluctuations are completely screened (fig.~\ref{fig:theory_1} e) and f)). 
In this case, neutral bright excitons can freely move and bind efficiently with an increased number of free holes present in the system. Only at this high $p$-doping does the positive bright trion become the dominant emission line in the emission spectrum. The Fermi energy of the hole gas is then $\sim 20~\mathrm{meV}$ in comparison with  $\sim 29~\mathrm{meV}$ predicted for the transition from trion to polaron states \cite{PhysRevB.95.035417}. 

As shown in the supplementary information, the model developed here qualitatively reproduces the intensity trends observed in our experimental data (SI, fig.~7).

The evolution of emission energies for excitonic complexes in the $p$-doped WSe$_2$ ML (fig.~\ref{fig:161310_gate_dependence_2} b)) can be explained in the framework of our model. The monotonous decrease of the bright and dark trion emission energies with increasing hole concentration can be attributed to an increased Coulomb screening from free holes in the system. Congruent energy shifts in the X$^+$ and X$^+_\mathrm{d}$ lines confirm their similar particle configuration. The observed minimum in the X$^0$ emission energy can be attributed to a Stark shift associated with the localization-induced charge separation. The screening of the potential fluctuations reduces the Stark shift, leading to an increased X$^0$ emission energy at moderate hole concentrations. Once the potential fluctuations are fully screened at approximately $-1~\mathrm{V}$, the emission energy starts to decrease due to the increased screening of the inter-excitonic Coulomb interactions (SI, fig.~5). The fact that the lowest X$^0$ emission energy is observed at a finite hole doping points to a hole-induced activation of the potential fluctuations, e.g.~via charged impurities (SI, fig.~3).

The linewidths (fig.~\ref{fig:161310_gate_dependence_2} c)) and shapes (SI, fig.~4) of the observed excitonic complexes support our theory (see SI for an extended discussion). At low hole concentrations, we deduce a decrease (increase) of the X$^0$ (X$^+$) lifetime with increasing $p$-doping, while the X$^+_\mathrm{d}$ lifetime remains unchanged in this regime. The decrease of the X$^0$ lifetime can be attributed to its increased radiative recombination rate and the more efficient transformation into positive bright trions. The extended X$^+$ lifetime is a result of the reduced scattering into positive dark trions, which results from the delocalization of positive bright trions. 

In conclusion, we have presented detailed gate-tunable photoluminescence data from a high-quality WSe$_2$ ML. We have shown that the photoluminescence intensity of neutral excitons is maximized at a finite hole doping level. Low doping levels favor the formation and recombination of positive dark trions, indicating that the Stark effect and, hence, charge noise effects, play a more important role in TMD-MLs than suggested by experiments with pure out-of-plane electric fields \cite{Roch2018}. Formation of positive bright trions requires higher  doping levels and is much less efficient than in electron doped system. We have explained this peculiar excitonic interplay in the hole-doped regime with a microscopic model based on exciton localization and doping dependent bright trion formation rates.  Although our theory fully describes the observed effects in a qualitative way, further experimental studies with a high resolution in the time domain need to be conducted, and quantitative theories developed that include both bright trions or attractive Fermi-polarons in the system. 
Our findings can be readily transferred to similar systems, such as WS$_2$ MLs and more broadly are an important building block in the understanding of exciton physics in two-dimensional systems. 
\\
\\
All samples were fabricated in the Helmholtz Nano Facility at Forschungszentrum J\"ulich \cite{hnf}. Growth of hexagonal boron-nitride crystals was supported by the Elemental Strategy Initiative conducted by the MEXT, Japan and the CREST (JPMJCR15F3), JST. The authors are grateful for fruitful discussions with Hanan Dery.

\bibliography{references}

\newpage
\clearpage
\newpage

\onecolumngrid
\noindent \begin{center}\textbf{\large Supplementary Information:\\Interplay of Excitonic Complexes in $p$-Doped WSe$_2$ Monolayers}\end{center}
\twocolumngrid

\setcounter{figure}{0}
\setcounter{page}{1}

\section{I. Sample Fabrication}

\begin{figure}[b]
    \centering
    \includegraphics{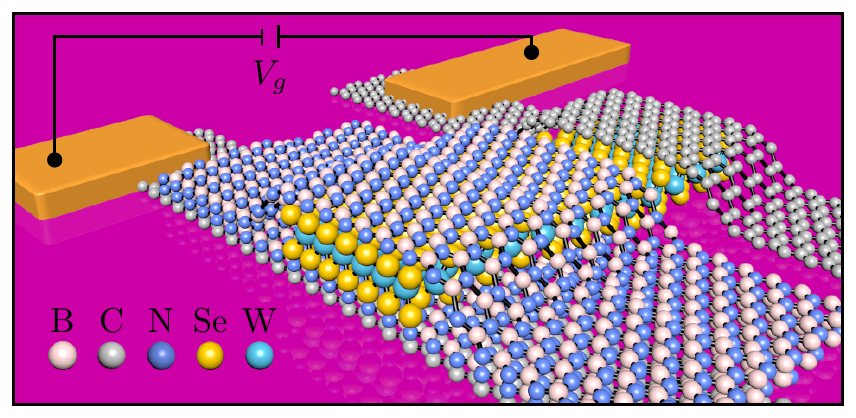}
    \caption{Sample design for gate-dependent photoluminescence measurements of a WSe$_2$ ML. The WSe$_2$ ML is encapsulated in two few-layer h-BN flakes and placed on a few-layer graphite back gate. Another few-layer graphite flake is used as an electrical contact to the WSe$_2$ ML. Both graphite flakes are contacted using metal contacts, allowing for the application of a gate voltage, $V_g$, between the graphite back gate and the WSe$_2$ ML.}
    \label{fig:sample_geometry}
\end{figure}

\begin{figure}[b]
    \centering
    \includegraphics{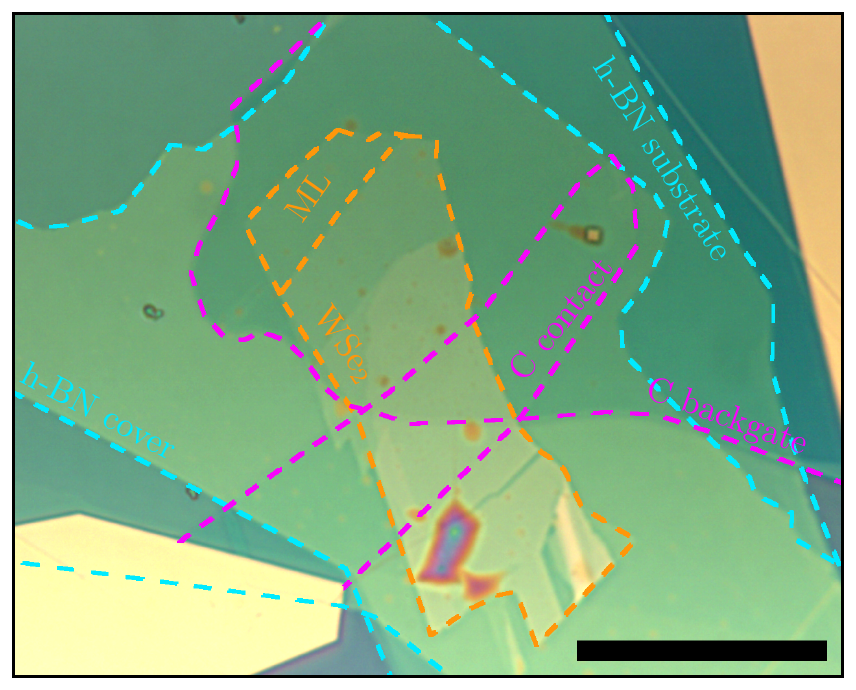}
    \caption{Optical micrograph of the gate-tunable WSe$_2$ ML device. The edges of the flakes involved in the device are marked by colored dashed lines and labeled. The metal contacts to the few-layer graphene flakes can be seen in the upper right and the lower left corners. The scale bar is $20~\mu\mathrm{m}$.}
    \label{fig:G26_micrograph}
\end{figure}

The high-quality WSe$_2$ ML device involves multiple van der Waals materials which form a stacked vertical heterostructure. A schematic diagram and an optical micrograph of the device are shown in figures \ref{fig:sample_geometry} and \ref{fig:G26_micrograph}, respectively. 

The WSe$_2$ ML is encapsulated within two thin flakes of h-BN to provide a high-quality substrate as well as a protection from the environment \cite{PhysRevX.7.021026, Wierzbowski2017, Ajayi_2017}. Furthermore, the lower h-BN flake with a thickness of $12~\mathrm{nm}$ acts as an insulator between the WSe$_2$ ML and an underlying back gate composed of a few-layer graphene flake. A second few-layer graphene flake is used as an electrical contact to the WSe$_2$ ML. 

The gate-tunable WSe$_2$ ML device was fabricated using mechanical exfoliation and stacking techniques, which are commonly used in the processing of van der Waals heterostructures \cite{doi:10.1063/1.4886096}. As a first step, bulk crystals of each material were cleaved on an adhesive wafer tape (\textit{ULTRON SYSTEMS 1007R-7.3}) and then exfoliated directly onto Si/SiO$_2$ chips. The chips were scanned for flakes with the right thickness, size, and shape. The cleanliness of each flake was confirmed using atomic force microscopy prior to the stacking.

Starting with the uppermost h-BN flake, the chosen flakes were picked up from the Si/SiO$_2$ chips one after another using a stamp consisting of a thick PDMS (\textit{SYLGARD 184}) film and a thin Polycarbonate (\textit{Sigma-Aldrich}) film. The final stack consisting of five flakes was then released on another Si/SiO$_2$ chip with predefined metal markers for electron beam lithography. An electron beam lithography process using Poly(methyl methacrylate) (PMMA) as a resist was performed in order to provide Ti/Au contacts to the few-layer graphene flakes.

\section{II. Optical Measurements}

All optical measurements were performed in a liquid helium cooled cold finger cryostat at $T=11~\mathrm{K}$. The sample was excited with a $1.88~\textrm{eV}$ continuous-wave laser, which was focused to a spot size of approximately $1.6~\mu\mathrm{m}$ on the sample surface using an aspheric lens with a numerical aperture of $0.47$. The photoluminescence signal from the sample was collected through the same lens and analyzed using a silicon CCD at the output of a spectrometer. The excitation power density was set to $\sim5~\frac{\mu\mathrm{W}}{\mu\mathrm{m}^2}$. Using typical values for the absorbance \cite{PhysRevB.90.205422} we obtain a generation rate for excitons in the ML of $10^{19}~\mathrm{cm}^{-2}s$. When taking into account typical values for the dark exciton lifetime \cite{PhysRevB.96.155423} in WSe$_2$ MLs, we obtain a moderate exciton density of $10^9~\mathrm{cm}^{-2}$, which is much lower than exciton densities at which non-linear effects such as exciton-exciton annihilation occur \cite{PhysRevB.89.125427}.

The density of charge carriers within the WSe$_2$ ML was tuned by applying gate voltages between $-1~\mathrm{V}$ and $2~\mathrm{V}$ to the few-layer graphene back gate. The charge density calculated using the parallel capacitor model with a h-BN thickness of $12$~nm and a h-BN dielectric permittivity of $3.76$ \cite{Laturia2018} ranges from $p=3.1\cdot 10^{12}$~cm$^{-2}$ to $n=2.1\cdot 10^{12}$~cm$^{-2}$, the correction due to the quantum capacitance was neglected. With an effective electron mass of $m_e^* = 0.28~m_0$ \cite{Korm_nyos_2015}, the maximum Fermi energy of the electron gas is $17~\mathrm{meV}$ (compared with a negative trion binding energy of approximately $30~\mathrm{meV}$). The effective mass of holes in the lowest energy subband at the K-point is $m_h^* = 0.36~m_0$ \cite{Korm_nyos_2015} and the Fermi energy of the hole gas at $-1~\mathrm{V}$ is approximately $20~\mathrm{meV}$. Polaronic effects are expected to be significant at $E_\mathrm{F} >> \frac{4}{3}E_\mathrm{T}$, where $E_\mathrm{T} = 22~\mathrm{meV}$ is the binding energy of a positive trion \cite{PhysRevB.95.035417}. 

\section{III. Low-Energy Emission Lines from Impurities}

\begin{figure}[h!]
    \centering
    \includegraphics{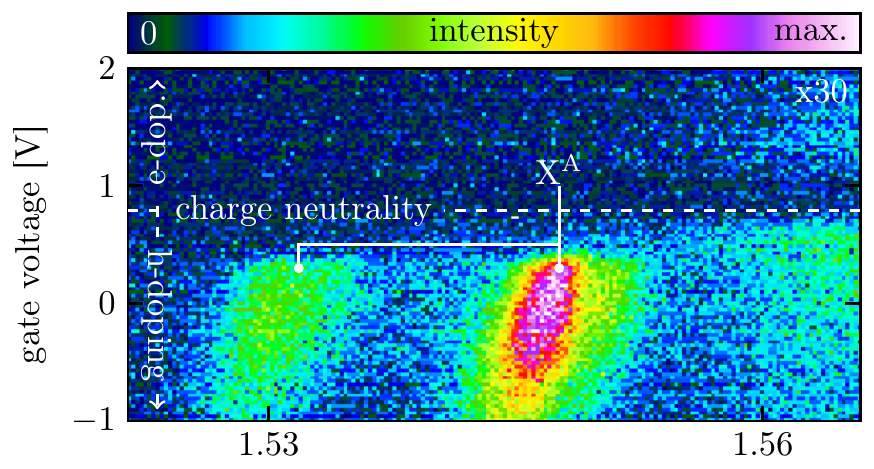}
    \caption{Low energy part of the photoluminescence spectrum of a gate-tunable WSe$_2$ ML. The shown data has been acquired together with the data shown in figure 1 of the main text. Two emission lines at low energy can be observed.}
    \label{fig:low_energy}
\end{figure}

The phenomenological model developed in the main text suggests charged impurities as a possible origin for exciton localization. Here, we present gate-dependent photoluminescence data of two emission lines at the low energy end of the energy spectrum measured in our experiments (fig.~\ref{fig:low_energy}). The energy of these lines ($1.54~\mathrm{eV}$ and $1.55~\mathrm{eV}$) agrees well with the theoretically predicted energy of acceptor bound excitons \cite{PhysRevLett.114.107401}. 

\newpage 

\section{IV. Photoluminescence Spectra at selected back gate voltages}

\begin{figure}[h!]
    \centering
    \includegraphics{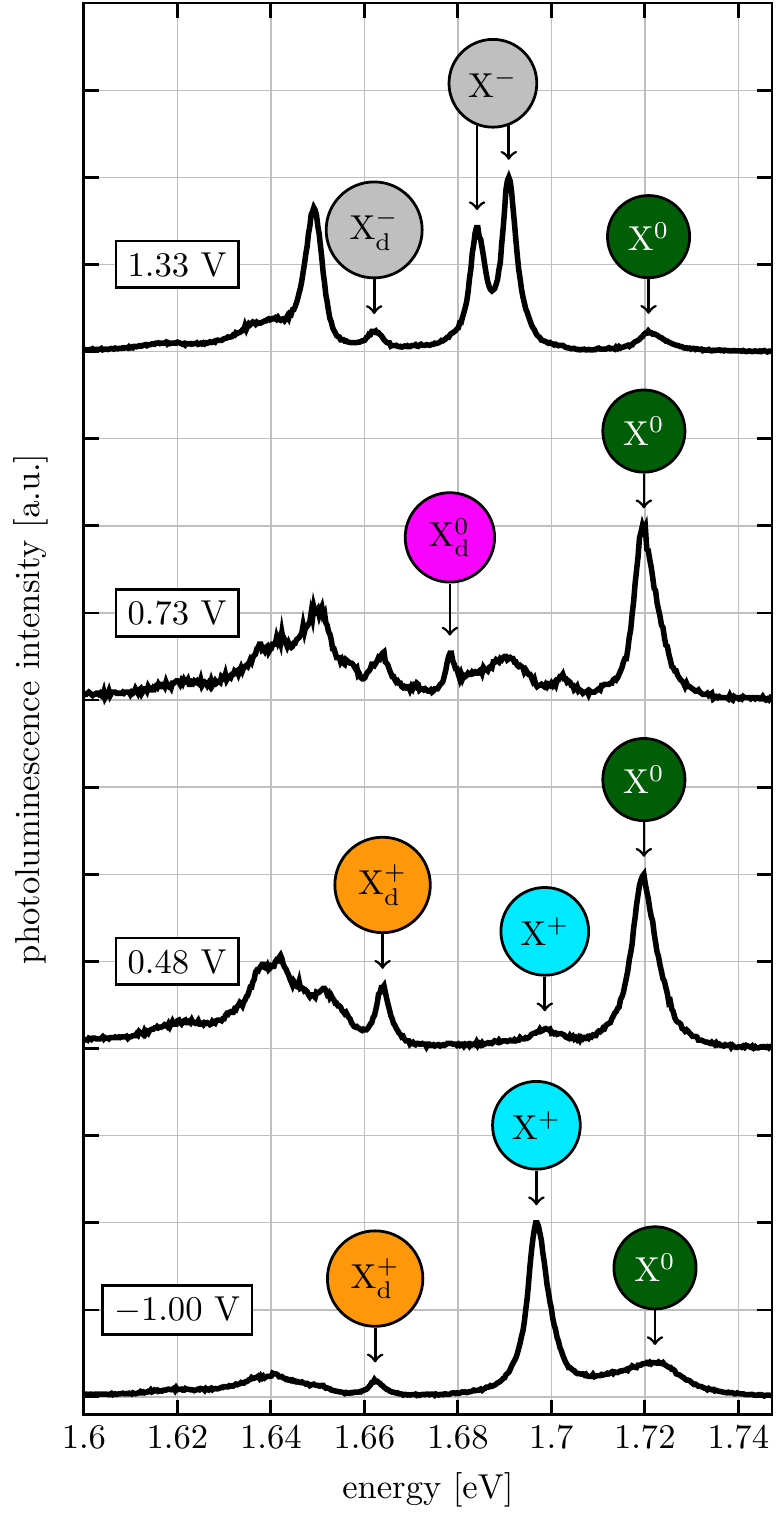}
    \caption{Photoluminescence spectra from a gate-tunable WSe$_2$ ML acquired at different back gate voltages. The shown spectra are extracted from the gate-dependent data shown in figure 1 of the main text and normalized for a better visualization. From top to bottom, the spectra show the photoluminescence signals from an electron-doped WSe$_2$ ML, a WSe$_2$ ML close to the neutrality point, a WSe$_2$ ML with a low hole doping and a WSe$_2$ ML with a high hole doping. The gate voltage for each spectrum is given next to the spectrum. The most important excitonic complexes are labeled.}
    \label{fig:spectra}
\end{figure}

\section{V. Discussion on Exciton Line Shapes}

The data discussed in the main text was analyzed by fitting each spectrum with a set of Voigt profiles. The Voigt profile is a function that describes well the line shape of many optical emitters. It can be described as the convolution of a Lorentzian and a Gaussian profile:

\begin{align*}
    V(E; I, E_0, \sigma, \gamma) &= I \int  G(\epsilon, \sigma) L(E-E_0-\epsilon, \gamma) d\epsilon\\
    G(E,\sigma) &= \frac{1}{\sigma \sqrt{2\pi}} e^{-\frac{-E^2}{\sigma\sqrt{2\pi}}} \\
    L(E,\gamma) &= \frac{\gamma}{\pi(E^2+\gamma^2)} \\
\end{align*}

\noindent Here, $V$, $G$, and $L$ are the Voigt, the Gaussian and Lorentzian profile, respectively; $I$ and $E_0$ denote the intensity and the peak energy of the emitter, respectively; $\sigma$ and $\gamma$ are the broadening parameters for the Gaussian and Lorentzian profile, respectively.

Several effects can contribute to the linewidth of an excitonic complex. While inhomogeneous broadening and slow energy noise lead to an increase of the Gaussian broadening, the decrease of a state's lifetime and fast energy noise result in an increase of the Lorentzian broadening. A detailed analysis of the fitted broadening parameters ($\sigma$ and $\gamma$) can, therefore, give further insights into the interplay of the studied excitonic complexes.

\begin{figure}[h!]
    \centering
    \includegraphics{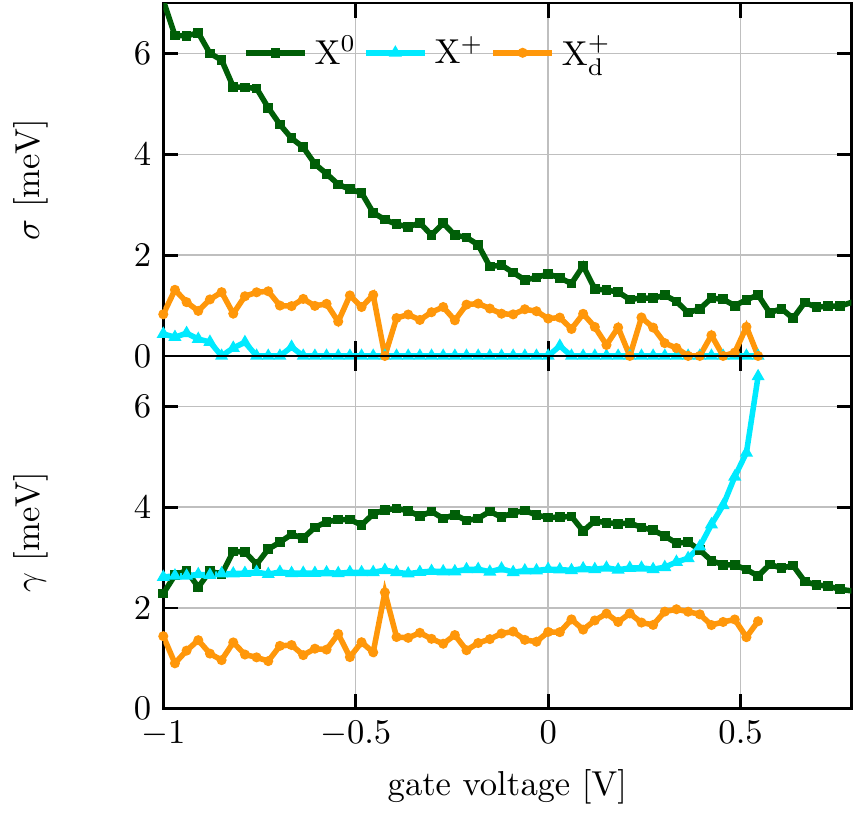}
    \caption{Line shape of the excitonic complexes, as a function of gate voltage. a) and b) show the broadening parameters for the Gaussian ($\sigma$) and Lorentzian ($\gamma$) profile, respectively, forming the Voigt profile.}
    \label{fig:161310_gate_dependence_3}
\end{figure}

\begin{figure}[t]
    \centering
    \includegraphics{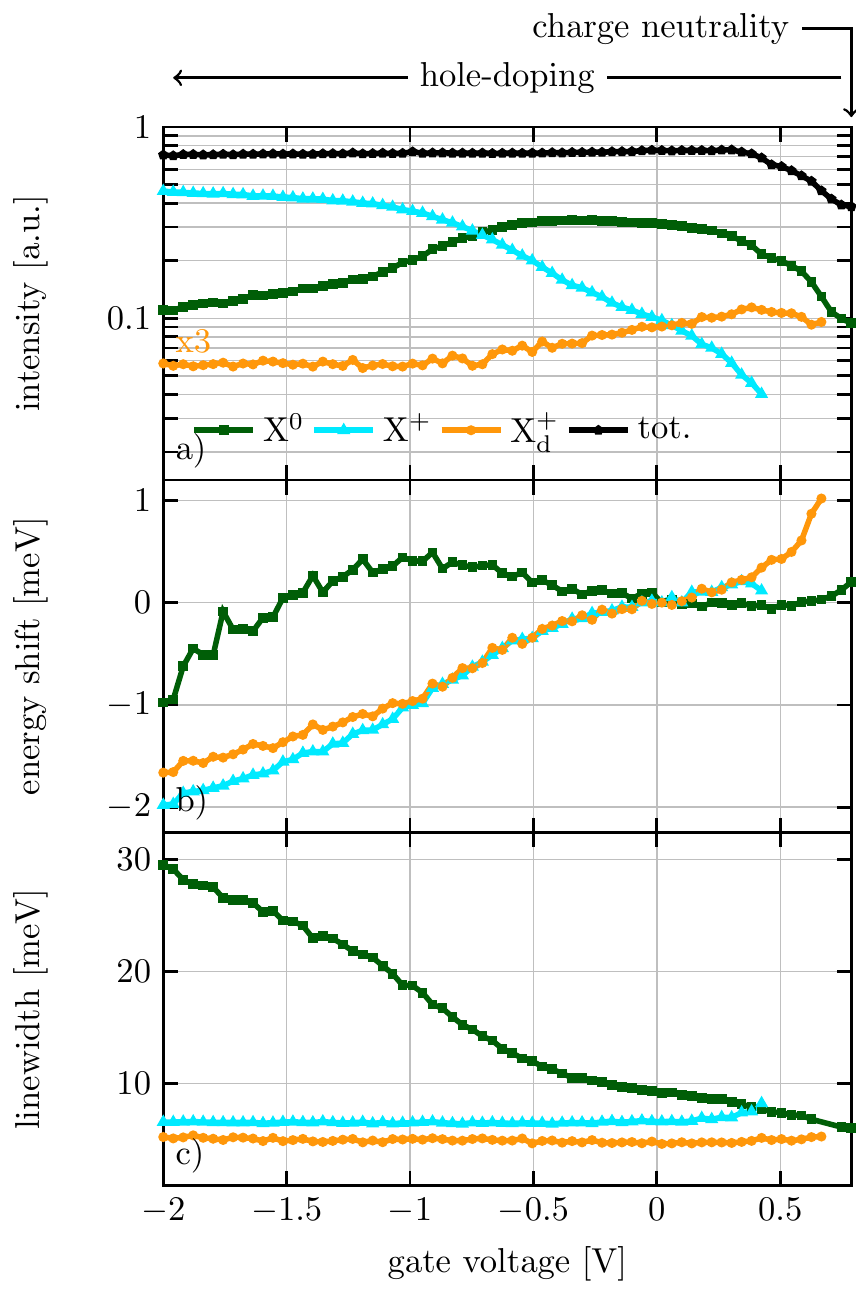}
    \caption{Same data as shown in figure 2 of the main text for another position on the same sample. The minimum applied gate voltage has been lowered to $-2~\mathrm{V}$. All other measurement parameters are kept the same. The measured linewidths of the excitonic complexes are slightly higher than in figure 2, which can be attributed to a stronger inhomogeneous broadening resulting from a larger disorder at this position on the sample. However, the data presented here confirms the conclusions from the main text.}
    \label{fig:180935_gate_dependence_2}
\end{figure}

Figure \ref{fig:161310_gate_dependence_3} shows $\sigma$ and $\gamma$ for the excitonic complexes discussed in the main text. It can be seen that the observed excitonic complexes are mostly (X$^0$, X$^+_\mathrm{d}$) or fully (X$^+$) Lorentzian at voltages between $0.5$~V and $0.35$~V, revealing insignificant contributions from inhomogeneous broadening. The simultaneous increase of the Lorentzian broadening and intensity of the X$^0$ line with an increasing $p$-doping close to the neutrality point indicates a negligible role of fast energy fluctuations and an increase of the X$^0$ radiative recombination rate. The fast decrease of the Lorentzian broadening of the X$^+$ line associated with the fast increase in its intensity points to an increase of the X$^+$ lifetime. Since both the X$^+_\mathrm{d}$ line shape and intensity remain constant at low $p$-concentrations, its lifetime does not change in this regime. At high $p$-concentrations, the X$^0$ line shows a strong contribution from inhomogeneous broadening, while the X$^+$ (X$^+_\mathrm{d}$) line shape remains unchanged (shows a weaker inhomogeneous broadening). 

\section{VI. Data from Another Position on the Sample}

The data presented in the main text is representative of the measured sample. Here, we show data from another position of this sample (fig.~\ref{fig:180935_gate_dependence_2}). For this measurement, the minimum gate voltage applied to the sample was extended to $-2~\mathrm{V}$.

In addition to our own experimental data, the phenomena discussed here have been observed in the published literature but have not been addressed \cite{PhysRevLett.123.027401,Li2019}.

\section{VII. Rate Equation System}

The transformation system for excitonic complexes in $p$-doped WSe$_2$ MLs shown in figure 3 of the main text can be translated into a rate equation system. This system consists of five differential equations which describe the change in the population of each excitonic complex, as well as the change in the hole population:

\begin{align*}
    \frac{dN_{\mathrm{X}^0}}{dt} &= g - \kappa_{\mathrm{X}^0}^{\mathrm{X}_\mathrm{d}^0} N_{\mathrm{X}^0} - \kappa_{\mathrm{X}^0}^{\mathrm{X}^+} N_{\mathrm{X}^0} p - \frac{N_{\mathrm{X}^0}}{\tau_{\mathrm{X}^0}} \\
    \frac{dN_{\mathrm{X}_\mathrm{d}^0}}{dt} &= \kappa_{\mathrm{X}^0}^{\mathrm{X}_\mathrm{d}^0} N_{\mathrm{X}^0} - \kappa_{\mathrm{X}_\mathrm{d}^0}^{\mathrm{X}_\mathrm{d}^+} N_{\mathrm{X}_\mathrm{d}^0} p - \frac{N_{\mathrm{X}_\mathrm{d}^0}}{\tau_{\mathrm{X}_\mathrm{d}^0}} \\
    \frac{dN_{\mathrm{X}^+}}{dt} &= \kappa_{\mathrm{X}^0}^{\mathrm{X}^+} N_{\mathrm{X}^0} p - \kappa_{\mathrm{X}^+}^{\mathrm{X}_\mathrm{d}^+} N_{\mathrm{X}^+} - \frac{N_{\mathrm{X}^+}}{\tau_{\mathrm{X}^+}} \\
    \frac{dN_{\mathrm{X}_\mathrm{d}^+}}{dt} &= \kappa_{\mathrm{X}_\mathrm{d}^0}^{\mathrm{X}_\mathrm{d}^+} N_{\mathrm{X}_\mathrm{d}^0} p + \kappa_{\mathrm{X}^+}^{\mathrm{X}_\mathrm{d}^+} N_{\mathrm{X}^+} - \frac{N_{\mathrm{X}_\mathrm{d}^+}}{\tau_{\mathrm{X}_\mathrm{d}^+}} \\
    \frac{dp}{dt} &= \frac{N_{\mathrm{X}^+}}{\tau_{\mathrm{X}^+}} + \frac{N_{\mathrm{X}_\mathrm{d}^+}}{\tau_{\mathrm{X}_\mathrm{d}^+}} - \kappa_{\mathrm{X}^0}^{\mathrm{X}^+} N_{\mathrm{X}^0} p - \kappa_{\mathrm{X}_\mathrm{d}^0}^{\mathrm{X}_\mathrm{d}^+} N_{\mathrm{X}_\mathrm{d}^0} p
\end{align*}

\noindent Here, $p$ and $N_x$ denote the hole population and the population of the excitonic complex $x$, respectively; $g$ is the generation rate for neutral bright excitons; $\kappa_i^f$ represents the transformation rate from the excitonic complex $i$ to the excitonic complex $f$, and $\tau_x$ is the lifetime of the excitonic complex $x$ taking into account both radiative and non-radiative recombination. From the exciton populations, the measured photoluminescence intensities, $I_x$, can be calculated:

\begin{align*}
    I_{\mathrm{X}^0} &= \frac{N_{\mathrm{X}^0}}{\tau_{\mathrm{X}_\mathrm{d}^0}^{\mathrm{rad.}}}\\
    I_{\mathrm{X}_\mathrm{d}^0} &= \frac{N_{\mathrm{X}_\mathrm{d}^0}}{\tau_{\mathrm{X}_\mathrm{d}^0}^{\mathrm{rad.}}}d\\
    I_{\mathrm{X}^+} &= \frac{N_{\mathrm{X}^+}}{\tau_{\mathrm{X}^+}^{\mathrm{rad.}}}\\
    I_{\mathrm{X}_\mathrm{d}^+} &= \frac{N_{\mathrm{X}_\mathrm{d}^+}}{\tau_{\mathrm{X}_\mathrm{d}^+}^{\mathrm{rad.}}}d\\
\end{align*}

\noindent with $\tau_x^{\mathrm{rad.}}$ and $d$ being the radiative lifetime of the excitonic complex $x$ and a geometrical factor accounting for differences in the collection efficiencies of bright and dark states due to different radiation patterns, respectively.

We consider the simple case of a perfect semiconductor and neglect all non-radiative recombination channels. For this case, the parameters of the rate equation system can be approximated as

\begin{align*}
    g &= \mathrm{const.}\\
    \kappa_{\mathrm{X}^0}^{\mathrm{X}_\mathrm{d}^0} &= \kappa_{\mathrm{X}^+}^{\mathrm{X}_\mathrm{d}^+} = \mathrm{const.}\\
    \kappa_{\mathrm{X}^0}^{\mathrm{X}^+} &= \kappa_{\mathrm{X}_\mathrm{d}^0}^{\mathrm{X}_\mathrm{d}^+} =  \mathrm{const.}\\
    \tau_{\mathrm{X}^0} &= \tau_{\mathrm{X}^0}^{\mathrm{rad.}} = \mathrm{const.}\\
    \tau_{\mathrm{X}^+} &= \tau_{\mathrm{X}^+}^{\mathrm{rad.}} = \mathrm{const.}\\
    \tau_{\mathrm{X}_\mathrm{d}^0} &= \tau_{\mathrm{X}_\mathrm{d}^0}^{\mathrm{rad.}} = \mathrm{const.}\\ \tau_{\mathrm{X}_\mathrm{d}^+} &= \tau_{\mathrm{X}_\mathrm{d}^+}^{\mathrm{rad.}} = \mathrm{const.}\\
\end{align*}

\begin{figure}
    \centering
    \includegraphics{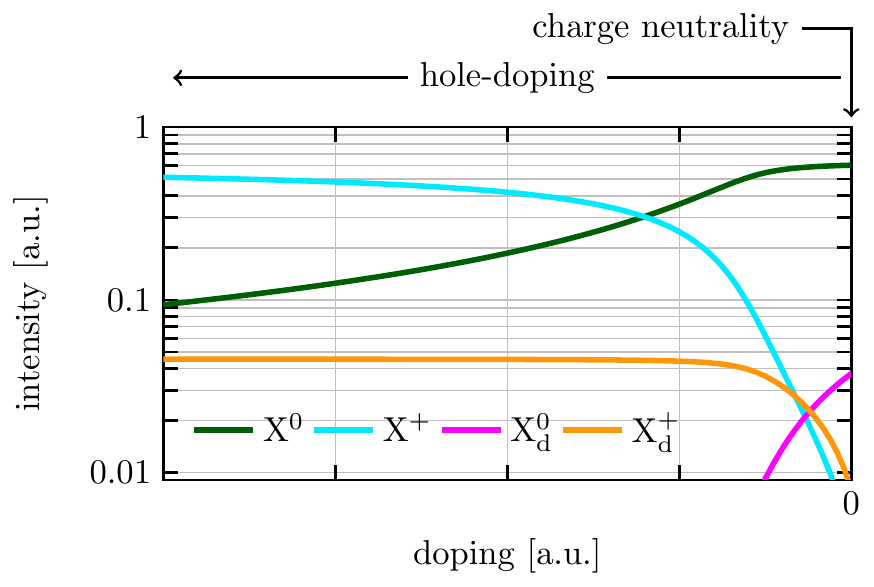}
    \caption{Calculated photoluminescence intensities for excitonic complexes in the $p$-doped regime of a perfect WSe$_2$ ML. It can be seen that the highest intensity for neutral excitonic complexes is obtained at the charge neutrality point. The intensity of neutral (positively charged) excitonic complexes decreases (increases) as a function of hole doping.}
    \label{fig:simulation}
\end{figure}

\noindent Using these assumptions, the rate equation system has been solved numerically and the calculated intensities for the excitonic complexes in $p$-doped WSe$_2$ ML are shown in figure \ref{fig:simulation}. 

The transformation rates and lifetimes can be adapted to the microscopic model developed in the main text. In this case, the transformation rate from neutral bright excitons to positive bright trions is expected to increase as a function of the screening from charged particles in the ML. Furthermore, it increases as a function of the hole concentration, which act as energy absorbers in the process of bright trion formation. We, therefore, assume 

\begin{equation*}
    \kappa_{\mathrm{X}^0}^{\mathrm{X}^+} \propto \begin{cases}
q p & \mathrm{for}~0 < q < q_s\\
p & \mathrm{for}~q > q_s
\end{cases}\\
\end{equation*}

\noindent with $q = p + N_{\mathrm{X}^+} + N_{\mathrm{X}^+_\mathrm{d}}$ being the total charge in the system and $q_s$ being the charge which fully screens the potential fluctuations. At the same time, the radiative {\parfillskip0pt\par} \newpage

\begin{figure}[!t]
    \centering
    \includegraphics{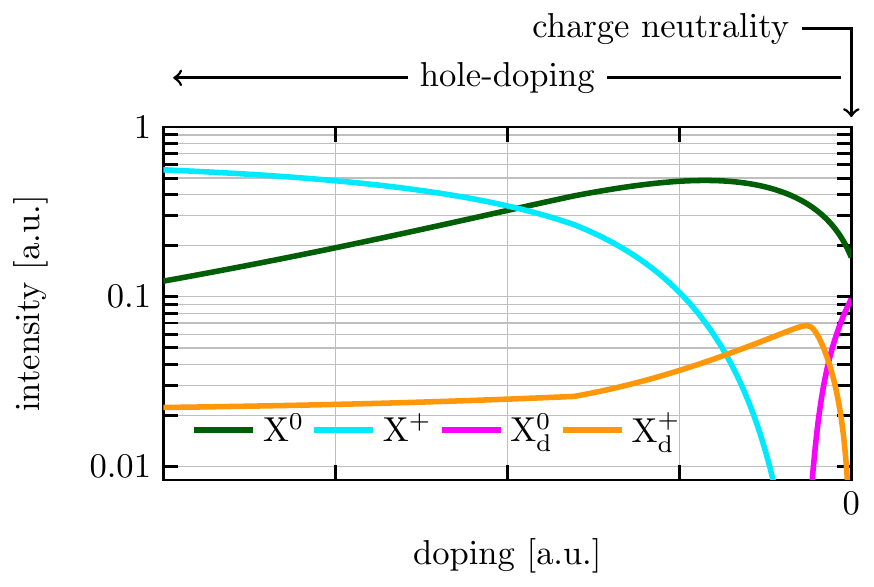}
    \caption{Calculated photoluminescence intensities for excitonic complexes in the $p$-doped regime of a WSe$_2$ ML assuming potential fluctuations, which are gradually screened with increasing hole doping, and taking into account the role of free holes as energy absorbers in the formation process of positive bright trions. The calculation agrees qualitatively with the presented experimental data (main text fig.~2, SI fig.~\ref{fig:180935_gate_dependence_2}). Neutral bright excitons show a maximum intensity at moderate hole doping, whereas neutral dark excitons vanish quickly as holes are being added to the system. Positive bright and dark trions show opposite behaviours, as positive bright (dark) trions increase (decrease) with increasing hole density.}
    \label{fig:simulation_2}
\end{figure}

~

\newpage

\noindent lifetime of the neutral bright excitons is expected to decrease with increasing screening from charge carriers in the system. Here, we assume

\begin{equation*}
    \tau_{\mathrm{X}^0} \begin{cases}
\propto (1+cq)^{-1} & \mathrm{for}~0 < q < q_s\\
=\mathrm{const.} & \mathrm{for}~q > q_s
\end{cases} \\
\end{equation*}
\noindent with $c$ being a constant. The modified rate equation system for the case of gradually screened potential fluctuations in the sample was solved numerically and the calculated intensities for the excitonic complexes in $p$-doped WSe$_2$ ML are shown in figure \ref{fig:simulation_2}. It can be seen that the trends in the measured intensities of the excitonic complexes are qualitatively reproduced by this simple model. It should be noted that the specific form of the functions representing $\kappa_{\mathrm{X}^0}^{\mathrm{X}^+}$ and $\tau_{\mathrm{X}^0}$ does not change the trends shown in the plot in figure 7 as long as $\kappa_{\mathrm{X}^0}^{\mathrm{X}^+}$ ($\tau_{\mathrm{X}^0}$) increases (decreases) as a function of hole concentration. The exact determination of the $p$-dependence of these quantities will be a subject of further studies.

\clearpage

\end{document}

%% file: author_list.tex
\author{Sven Borghardt} \affiliation{Peter Gr\"{u}nberg Institute 9, Forschungszentrum J\"{u}lich, 52425 J\"{u}lich, Germany}\affiliation{Department of Physics, RWTH Aachen University, 52074 Aachen, Germany}
\author{Jhih-Sian Tu} \affiliation{Helmholtz Nano Facility, Forschungszentrum J\"{u}lich, 52425 J\"{u}lich, Germany}
\author{Takashi Taniguchi}
\author{Kenji Watanabe}\affiliation{National Institute for Materials Science, Tsukuba 305-0047 Ibaraki, Japan}
\author{Beata E.~Kardyna\l} \affiliation{Peter Gr\"{u}nberg Institute 9, Forschungszentrum J\"{u}lich, 52425 J\"{u}lich, Germany}\affiliation{Department of Physics, RWTH Aachen University, 52074 Aachen, Germany}